\documentclass[a4paper,12pt]{article}
\pdfoutput=1
\usepackage[utf8]{inputenc}
\usepackage{amsmath, amssymb, amsfonts}
\usepackage{geometry}
\usepackage{amsmath}
\usepackage{graphicx,rotating,hyperref,slashed,amsmath,xcolor,amssymb,amsfonts,colortbl,cite, subfigure,float}

\usepackage{graphicx,rotating,hyperref,slashed,xcolor,amsmath,amssymb,amsfonts,colortbl,subfigure,float,mathrsfs,mathalpha}
\usepackage{graphics}
\usepackage{comment} 

\hypersetup{colorlinks,bookmarksopen,bookmarksnumbered,
linkcolor=blue,pdfstartview=FitH,urlcolor=blue,citecolor=blue}
\numberwithin{equation}{section}

\usepackage[utf8]{inputenc}
\usepackage[T1]{fontenc}
\usepackage{lmodern}
\usepackage{geometry}
\usepackage{hyperref}
\usepackage{enumitem}
\usepackage{setspace}

 \usepackage[utf8]{inputenc}
\usepackage[T1]{fontenc}
\usepackage[french,english]{babel}
\usepackage{csquotes}
\usepackage{setspace}
\usepackage{geometry}

\usepackage{graphicx,rotating,hyperref,slashed,amsmath,xcolor,amssymb,amsfonts,colortbl,cite} 
\makeatletter
\hypersetup{colorlinks,bookmarksopen,bookmarksnumbered,
linkcolor=blus,pdfstartview=FitH,urlcolor=rossos,citecolor=verde}
\allowdisplaybreaks	 
\numberwithin{equation}{section}

\hypersetup{%
    ,urlcolor=black
    ,citecolor=black
    ,linkcolor=black
    }
\usepackage{mciteplus}

\AtBeginDocument{
  \hypersetup{
    urlcolor=blue,
    citecolor=blue,
    linkcolor=black,
  }%
}

\def\beq{\begin{equation}}
\def\eeq{\end{equation}}

\renewcommand\[{\left[}

\newcommand\ees{\end{eqnarray}}
\newcommand\bees{\begin{eqnarray}}

\def\bea{\begin{eqnarray}}
\def\eea{\end{eqnarray}}

\def\0{{\boldsymbol 0}}

\def\lsim{\mathrel{\rlap{\lower3pt\hbox{\hskip0pt$\sim$}}
   \raise1pt\hbox{$<$}}}         
\def\gsim{\mathrel{\rlap{\lower4pt\hbox{\hskip1pt$\sim$}}
   \raise1pt\hbox{$>$}}}         

 \newcommand{\sfootnote}[1]{}
\definecolor{bluc}{cmyk}{1,1,0,0.1}
\definecolor{rossoCP3}{cmyk}{0,.88,.77,.40}
\definecolor{rosso}{cmyk}{0,1,1,0.4}
\definecolor{rossos}{cmyk}{0,1,1,0.55}
\definecolor{rossoc}{cmyk}{0,1,1,0.2}
\definecolor{verdes}{cmyk}{0.92,0,0.59,0.4}

\hypersetup{colorlinks, bookmarksopen, bookmarksnumbered,
citecolor=verdes, linkcolor=bluc, pdfstartview=FitH, urlcolor=rossos}

\usepackage{multicol}
\usepackage{color}
\definecolor{rosso}{cmyk}{0,1,1,0.4}
\definecolor{rossos}{cmyk}{0,1,1,0.55}
\definecolor{rossoc}{cmyk}{0,1,1,0.2}
\definecolor{blu}{cmyk}{1,1,0,0.3}
\definecolor{blus}{cmyk}{1,1,0,0.6}
\definecolor{bluc}{cmyk}{1,1,0,0.1}
\definecolor{verde}{cmyk}{0.92,0,0.59,0.25}
\definecolor{verdec}{cmyk}{0.92,0,0.59,0.15}
\definecolor{verdes}{cmyk}{0.92,0,0.59,0.4}

\skip\@mpfootins = \skip\footins
\def\circa#1{\,\raise.3ex\hbox{$#1$\kern-.75em\lower1ex\hbox{$\sim$}}\,}

\newcommand{\be}{\begin{equation}}
\newcommand{\ee}{\end{equation}}
\newfam\rsfsfam
\def\mathscr#1{{\fam\rsfsfam\relax#1}}

\def\circa#1{\,\raise.3ex\hbox{$#1$\kern-.75em\lower1ex\hbox{$\sim$}}\,}
\makeatletter

\def\hhref#1{\href{http://arxiv.org/abs/#1}{arXiv:#1}} 

\newcommand{\doi}[1]{\href{http://dx.doi.org/#1}{[doi]}}

\def\hhref#1{\href{http://arxiv.org/abs/#1}{arXiv:#1}} 
 
\def\art{\@ifnextchar[{\eart}{\oart}}
\def\eart[#1]#2#3#4#5#6{{\rm #2}, {\em #3 \bf #4} {\rm (#6) #5} ({\em #1})}

\def\article{\@ifnextchar[{\earticle}{\oarticle}}
\def\oarticle#1#2#3#4#5#6{{\rm #1}, {\em ``#6''}, {\rm #2 #3 (#5) #4}}
\def\earticle[#1]#2#3#4#5#6#7{{\rm #2}, {\em ``#7''}, {\rm #3 #4 (#6) #5}  [\hhref{#1}]}
\def\hepart[#1]#2{{\rm #2, \em#1}}
\def\heparticle[#1]#2#3{#2, {\em ``#3''} [\hhref{#1}]}

\newcounter{alphaequation}[equation]
\def\thealphaequation{\theequation\hbox to
0.6em{\hfil\alph{alphaequation}\hfil}}
\def\eqnsystem#1{
\def\@eqnnum{{\rm (\thealphaequation)}}
\def\@@eqncr{\let\@tempa\relax \ifcase\@eqcnt \def\@tempa{& & &} \or
  \def\@tempa{& &}\or \def\@tempa{&}\fi\@tempa
  \if@eqnsw\@eqnnum\refstepcounter{alphaequation}\fi
\global\@eqnswtrue\global\@eqcnt=0\cr}
\refstepcounter{equation} \let\@currentlabel\theequation \def\@tempb{#1}
\ifx\@tempb\empty\else\label{#1}\fi
\refstepcounter{alphaequation}
\let\@currentlabel\thealphaequation
\global\@eqnswtrue\global\@eqcnt=0 \tabskip\@centering\let\\=\@eqncr
$$\halign to \displaywidth\bgroup \@eqnsel\hskip\@centering
$\displaystyle\tabskip\z@{##}$&\global\@eqcnt\@ne
\hskip2\arraycolsep\hfil${##}$\hfil& \global\@eqcnt\tw@\hskip2\arraycolsep
$\displaystyle\tabskip\z@{##}$\hfil
\tabskip\@centering&\llap{##}\tabskip\z@\cr}
\def\endeqnsystem{\@@eqncr\egroup$$\global\@ignoretrue} \makeatother

\definecolor{fiorentina}{rgb}{.5,0,.5}

\begin{document}

\setcounter{page}{1} \baselineskip=15.5pt \thispagestyle{empty}

\vspace{0.8cm}
\begin{center}

{\fontsize{16}{28}\selectfont  \sffamily \bfseries {
Transient Parity Violation during Inflation: \\Implications for PTA Gravitational Waves
}}

\end{center}

\vspace{0.2cm}

\begin{center}
{\fontsize{12}{30}\selectfont  Gianmassimo Tasinato$^{1,2}$ } 
\end{center}

\begin{center}

\vskip 8pt
\textsl{$^{1}$ Physics Department, Swansea University, SA28PP, United Kingdom}\\
\textsl{$^{2}$ Dipartimento di Fisica e Astronomia, Universit\`a di Bologna,  Italy}\\
\textsl{\texttt{email}: g.tasinato2208 at gmail.com }\\
\vskip 7pt

\end{center}

\smallskip

\begin{abstract}
\noindent
{\small
We investigate the consequences of a transient phase of enhanced parity violation during inflation. Modeling this phase through a time-localized Chern--Simons-like coupling, we show that it amplifies primordial gravitational waves at small scales, producing a robust spectral shape with a blue growth of effective slope $n_T \simeq 2$, largely insensitive to microscopic details. 
This prediction lies in the range explored by recent pulsar timing array (PTA) analyses under cosmological power-law interpretations, while differing from the canonical supermassive black hole binary expectation. Our framework thus provides a predictive cosmological template to benchmark astrophysical versus primordial origins of the signal, consistent with cosmic microwave background bounds.
The signal also exhibits large linear polarization and non-trivial Stokes correlations, corresponding to an almost maximally phase-coherent helicity state. Such features are difficult to realize in classical stochastic backgrounds, and their detection would provide circumstantial evidence for a primordial, coherently generated origin of the gravitational-wave background.
}
\end{abstract}

\section{Introduction}

Cosmic inflation \cite{Starobinsky:1980te,Guth:1980zm,Linde:1981mu} provides a compelling framework for the physics of the early universe, strongly supported by large-scale cosmological observations. Among its most robust predictions is the generation of a stochastic background of primordial gravitational waves (GW) \cite{Starobinsky:1979ty}. 
So far, experimental efforts to probe inflationary GW have primarily focused on their imprints in the cosmic microwave background (CMB); see e.g.~\cite{Kamionkowski:2015yta} for a review. 
However,  
 it is both timely and well motivated to explore whether inflation can also generate GW signals directly accessible to present or near-future gravitational-wave observatories (see e.g.~\cite{Bartolo:2016ami}). In particular, can inflation naturally predict a GW signal with amplitude and spectral slope compatible with recent pulsar timing array (PTA) measurements~\cite{NANOGrav:2023gor,Reardon:2023gzh,Xu:2023wog,EPTA:2023fyk}?

\smallskip

In this work, we address this question by investigating the consequences of a transient phase during inflation in which parity-violating (PV) effects are significantly enhanced. We adopt a phenomenological framework in which interactions with an additional sector are localized in time, reflecting their short-lived nature. Concretely, we consider a coupling inspired by gravitational Chern--Simons interactions \cite{Jackiw:2003pm}, modulated by a time-dependent function that becomes active only for a brief interval during inflation.

Within this setup, we show that a transient PV phase can lead to an efficient amplification of primordial GW at small scales.  A key result of our analysis is that the GW spectrum exhibits a {\it universal} behavior: its shape is well approximated by a power-law with a  growth rate $n_T \simeq 2$, largely independent of the microscopic details of the model. We interpret this behaviour as due to the
would-be tensor decaying mode, which becomes active and plays a role in this scenario. 
This prediction lies within the range obtained in PTA analyses when the signal is interpreted using cosmological power-law templates (see e.g.~\cite{NANOGrav:2023hvm,EPTA:2023xxk,Figueroa:2023zhu,Vagnozzi:2023lwo,Ellis:2023oxs}). Hence our mechanism provides a concrete realization of a predictive cosmological template with minimal parametric freedom. 
  On the other hand, the canonical astrophysical interpretation in terms of supermassive black hole binaries predicts $\Omega_{\rm GW} \propto f^{2/3}$, although environmental effects, eccentricity, and population properties can modify this expectation. At present,   both interpretations remain viable. 

\smallskip

In addition,
the resulting stochastic background is characterized not only by its amplitude, but also by distinctive polarization features, including both circular and linear components.
 Our scenario 
predicts specific correlations between the GW intensity and the associated Stokes parameters, thereby enhancing its observational testability. In particular, the presence of sizable linear polarization emerges as a distinctive signature of the transient PV dynamics, offering a further difference with respect to astrophysical predictions.
Moreover, we propose that our findings can be interpreted  as providing circumstantial indications of a  coherently generated origin of the GW background.


\smallskip

The structure of this paper is as follows:

\smallskip
\noindent
$\bullet$  In Section~\ref{sec_setup}, we introduce our setup for transient parity-violating interactions during inflation, placing it in the context of the existing literature. We derive fully analytic solutions for the PV mode functions, and we demonstrate that the resulting GW spectrum exhibits universal features, insensitive to the detailed dynamics of the transient phase.

\smallskip
\noindent
$\bullet$
 In Section~\ref{sec_pheno}, we explore the phenomenological implications for PTA experiments. We show that the universal spectral slope identified in Section~\ref{sec_setup} 
is able reproduce the preferred amplitude and effective slope
for current PTA data, while remaining consistent with CMB bounds on  tensor modes. We also compute the scale dependence of the GW Stokes parameters, highlighting the presence of a significant amount of  GW linear polarization which provides  a distinctive signature of the scenario.

\smallskip
\noindent
$\bullet$ In Section~\ref{sec_quantum}, we show that our scenario predicts well-defined correlations in GW helicity space, corresponding to a phase-coherent polarization pattern. Such a configuration is difficult to realize with conventional astrophysical sources, and its detection would provide suggestive evidence for a primordial, coherently generated origin of the observed GW signal.

\smallskip
\noindent
$\bullet$ We conclude in Section~\ref{sec_out} with a discussion of future directions and possible extensions of our research.

\smallskip
Throughout this work we make use of  natural units $c = \hbar = M_{\rm Pl} = 1$.

\section{Our setup}
\label{sec_setup}

A key motivation for our work is the current underdetermination of the origin of the PTA signal. While astrophysical models provide a well-motivated explanation, they are subject to significant uncertainties, and cosmological interpretations often rely on flexible or highly tuned spectral shapes. It is therefore important to develop predictive cosmological scenarios with robust and characteristic signatures. 

\smallskip 
For these reasons we introduce a novel mechanism for the amplification of the stochastic gravitational wave background produced by  inflation, yielding distinctive and testable predictions. Our approach leverages the impact of a brief transient phase characterized by enhanced parity-violating interactions. Remarkably, regardless of the details of the transition, this dynamics leads to a universal spectral slope for the growing part of the gravitational wave spectrum,
with interesting consequences for PTA measurements. Similar motivations, leading to different
realizations, drive the results developed   in \cite{Fu:2023aab,Niu:2023bsr}.

\subsection{Motivations} 

Several scenarios in the early universe can amplify the spectrum of gravitational waves (GW) to levels detectable by current and future experiments (see e.g. \cite{Caprini:2018mtu,LISACosmologyWorkingGroup:2022jok} for reviews). A well-studied possibility is the production of GW during the radiation-dominated era, sourced at second order by scalar fluctuations. In particular, enhanced scalar perturbations arising during ultra-slow-roll phases of inflation can lead to significant GW signals, often associated with primordial black hole (PBH) formation
 (see e.g. \cite{Domenech:2021ztg,Ozsoy:2023ryl,LISACosmologyWorkingGroup:2025vdz,Carr:2026hot} for reviews). 

While such scenarios can, in principle, reproduce the amplitude suggested by recent pulsar timing array (PTA) observations \cite{NANOGrav:2023gor,Reardon:2023gzh,Xu:2023wog,EPTA:2023fyk}, they are typically subject to stringent constraints from PBH overproduction \cite{NANOGrav:2023hvm}.  
Avoiding these bounds typically requires a degree of fine-tuning. This may be achieved, for example, by incorporating the effects of scalar non-Gaussianities,
or a non-standard equation of state in the early universe
(see, e.g.,~\cite{Liu:2023pau,Domenech:2024rks,Harigaya:2023pmw,Balaji:2023ehk,Franciolini:2023pbf,Liu:2023ymk,Liu:2023hpw}). 
This motivates the exploration of alternative mechanisms in which the GW spectrum is enhanced more directly during inflation or immediately after,
 breaking internal space-time symmetries \cite{Endlich:2012pz,Cannone:2014uqa,Bartolo:2015qvr}, 
introducing phases of rapid variations of GW speed \cite{Mylova:2018yap,Ozsoy:2019slf}, tuning the inflaton
potential \cite{Frosina:2023nxu}, 
 or through additional sectors besides the inflaton, which evade PBH constraints (see e.g. \cite{Ferrante:2023bgz,Gorji:2023sil}). 
An additional requirement is to construct scenarios in which the slope of the GW spectrum is not treated as a free parameter, but instead specific values are predicted {\it naturally} within the range favored by current observations \cite{NANOGrav:2023hvm,EPTA:2023xxk,Figueroa:2023zhu,Vagnozzi:2023lwo,Ellis:2023oxs}. In this work, we pursue this goal.

\smallskip

We focus on inflationary scenarios which violate parity, i.e. they predict a different
amount of left versus right GW polarizations produced during inflation \cite{Lue:1998mq,Alexander:2004us,Satoh:2007gn,Anber:2009ua,Contaldi:2008yz,Alexander:2009tp}. In our case, besides violating parity and switching on sizeable values for the GW Stokes
parameters, our mechanism also amplifies the GW intensity and  energy density.  
 Parity violating
interactions are well motivated in fundamental physics setup \cite{Lee:1956qn,Wu:1957my}, and if detected would
provide a compelling example of violation of fundamental discrete symmetries  during inflation (see also \cite{Cannone:2015rra}  for a description in terms of the effective
field theory approach).
The  starting point motivating our investigation is a Chern--Simons extension of gravity during inflation, in which a time-dependent scalar field $\phi$---for instance the inflaton---couples to the gravitational Chern--Simons term,
\begin{equation}
\label{exa_cs}
S = \int d^4 x \sqrt{-g} \, f(\phi)\, R_{\mu\nu\rho\sigma} \tilde{R}^{\mu\nu\rho\sigma} \, .
\end{equation}
Here $\tilde{R}^{\mu\nu\rho\sigma}$ denotes the dual of the Riemann tensor. Such interactions are known to induce parity-violating effects in the tensor (spin-2) sector, leading to a chiral GW background (see e.g. the review
\cite{Alexander:2009tp}).

In the simplest realization, the coupling function $f$ depends only on $\phi$. However, more general constructions based on the  effective field theory (EFT)    of inflation restricted to tensor modes (see e.g. \cite{Creminelli:2014wna})  
indicate that  derivatives of the scalar field also enter parity-violating operators \cite{Crisostomi:2017ugk,Gao:2019liu,Zhu:2022uoq}. Moreover, recent developments in open EFT for cosmology shown that parity-violating tensor structures of the same form naturally arise in a broader class of setups \cite{Lau:2024mqm,Salcedo:2025ezu}. All these facts motivate us to adopt a model-independent perspective in what follows.

We consider a Friedmann--Robertson--Walker background with mostly-plus signature, and denote by $a(\tau)$ the scale factor in conformal time. The Fourier modes of the tensor perturbations are written as $h_k^{(\lambda)}(\tau)$, where $k$ is the comoving momentum and $\lambda $ labels the helicity states,
which we  identify with the left and right polarizations of the
spin-2 modes:  $\lambda = \pm  = (L,R)$. Their evolution equation can be expressed as
\begin{equation}
u_k^{\lambda\,\prime\prime}(\tau) + \left[k^2 - \frac{z''(\tau)}{z(\tau)}\right] u_k^\lambda(\tau) = 0 \,.
\label{eq:cAk-helicalV2}
\end{equation}
where we denote with $u_k^{\lambda}(\tau)\,=\,z(\tau)\, h_k^{(\lambda)}(\tau)$
the canonically normalized version of the tensor mode. 
The function $z(\tau)$, often referred to as the pump field, encodes the effects of parity-violating interactions. For the specific case of Chern-Simons (CS) gravity, it is given by
\begin{equation}
\label{def_pump}
z(\tau) = a(\tau)\, \sqrt{1 - \lambda\, k\, \gamma(\tau)} \,.
\end{equation}
In the CS model we have 
\begin{equation}
\label{def_gaCS}
\gamma(\tau) = \frac{\partial_\phi f(\phi(\tau))}{a^2(\tau)} \,,
\end{equation}
where $f$ is the function appearing in Eq.~\eqref{exa_cs}. 
Notice that -- since it explicitly depends on the polarization $\lambda$ -- the pump field leads to PV effects in the evolution
of tensor modes.  

In order to encompass a wider class of theories---particularly within the EFT of inflation---we will treat Eq.~\eqref{eq:cAk-helicalV2} and the definition of $z(\tau)$ through
 Eq.~\eqref{def_pump}   as general parametrizations, without committing to a specific microscopic origin
of the function $\gamma(\tau)$ -- up to a specific example in Footnote~\ref{foot_exp}.

\medskip

Inflation is modeled as a phase of pure de Sitter evolution of constant Hubble parameter $H$ and scale factor $a(\tau) = -{1}/{(H \tau)}$; it occurs at negative conformal times  $\tau\le0$. We assume that the PV interactions  activate  only during a brief period of time. We parameterize the function $\gamma$ contained
in  Eq.~\eqref{def_pump} as piecewise evolution
\begin{eqnarray}
\gamma(\tau)&=&
\begin{cases}
0 \,, & \tau<\tau_1\,, \\[3pt]
\text{rapidly varying}\,, & \tau_1<\tau<\tau_2\,,\\[3pt]
0 \,, & \tau_2<\tau<0\,.
\end{cases}
\label{ans_gam}
\end{eqnarray}
The interval $(\tau_1,\tau_2)$ during inflation  is assumed to be short, 
\begin{equation}
\label{con_short}
\frac{\tau_2 - \tau_1}{\tau_1} \ll 1 \, .
\end{equation}
For simplicity we also assume that the (rapidly varying) function $\gamma(\tau)$ vanishes with its first
derivative when evaluated at  the times $\tau_1$ and $\tau_2$ -- so that it 
 smoothly connects to its vanishing value outside this time interval. 
A situation described by the Ansatz~\eqref{ans_gam} may arise, for instance, if during inflation a transient coupling to the environment generates local, parity-violating interactions of Chern--Simons type~\footnote{\label{foot_exp} One can, for instance, construct a function $f(\phi)$ in Eq.~\eqref{def_gaCS} with parametrically large second derivatives in a controlled interval. Assume $\phi(\tau)$ monotonically 
\be
f(\phi)=
\begin{cases}
\phi_1, & \phi\le \phi_1,\\[4pt]
\phi_1 + (\phi_2-\phi_1)\,\dfrac{\sigma^2}{\sigma^2+\epsilon(1-\sigma)^2}, & \phi_1\le \phi\le \phi_2,\\[6pt]
\phi_2, & \phi\ge \phi_2,
\end{cases}
\ee
with $\epsilon\ll 1$.
This function is continuous, has vanishing first derivatives at $\phi_{1,2}$, and features parametrically large second derivatives within the transition region, cf.~\eqref{exa_cs}. More generally, departures from slow-roll evolution can induce rapid variations in the inflaton velocity and in the slow-roll parameters, leading to localized features in the profile of the scalar function $\gamma$.}.
In what follows, we assume that $\gamma$ remains sufficiently bounded so that the argument of the square root in Eq.~\eqref{def_pump} stays positive over the range of scales relevant for PTA observations, thereby avoiding pathological behavior of the tensor modes. 
In this work, however, we adopt a deliberately phenomenological approach and do not commit to a specific microphysical realization of the Ansatz~\eqref{ans_gam}. Constructing explicit and theoretically consistent models that realize this dynamics is an important direction, which we leave for future investigation.

\smallskip

A brief but significant departure from the condition $\gamma = 0$ in the Ansatz~\eqref{ans_gam} leads to the excitation of tensor mode components  that would otherwise decay on superhorizon scales during inflation. (The so-called
decaying modes of  Eq.~\eqref{eq:cAk-helicalV2}.)
 Importantly, besides $\gamma(\tau)$, this effect also  depends on the helicity $\lambda$, so that the would-be decaying modes become dynamically relevant and contribute to an enhancement of the gravitational-wave signal.
This mechanism is closely analogous to what occurs for scalar perturbations in non-slow-roll phases of inflation, such as ultra-slow-roll scenarios often invoked in models of primordial black hole production (see, e.g., \cite{Ozsoy:2023ryl} for a review). In both cases, a transient violation of the attractor dynamics promotes the decaying mode into a growing one, leading to a substantial amplification of the corresponding fluctuations.
A key aspect of this phenomenon—both for scalar perturbations and for the parity-violating tensor modes considered here—is the emergence of {\it universal features} in the resulting GW spectrum, largely insensitive to the microphysical details of the underlying dynamics. This leads to robust and model-independent predictions for the gravitational-wave spectrum. It is precisely this degree of universality that makes our scenario particularly compelling, and which we now proceed to characterize.

\subsection{Solving the evolution equations}

Our goal is to determine a general solution of Eq.~\eqref{eq:cAk-helicalV2} for the mode function $\gamma$ satisfying the Ansatz~\eqref{ans_gam}. 

\smallskip

Before tackling the general case, it is useful to recall that for $\gamma=0$ the evolution equation during inflation admits the standard Bunch--Davies solution,
independent from the helicity index
\begin{equation}
h_k(\tau)=\frac{H}{\sqrt{2}\,k^{3/2}}\,e^{-i k \tau}
\left(1+ i k \tau \right).
\label{eq_stmf}
\end{equation}

\paragraph{General Ansatz.}
To incorporate the effects of a non-vanishing $\gamma$, we generalize Eq.~\eqref{eq_stmf} by introducing the following helicity-dependent Ansatz
for the canonically normalized mode functions $u_k^\lambda$ (which
enter in the mode equation)
\begin{equation}
\label{eq_ans2}
u^{\lambda}_k(\tau)=\frac{a(\tau)\,H}{\sqrt{2}\,k^{3/2}}\,e^{-i k \tau}
\Bigg[
1+ i k \tau + i\,\lambda\, \sum_{n=1}^{\infty} (i k \tau_0)^n \Gamma_n(\tau)
\Bigg],
\end{equation}
where $\lambda=\pm1$ labels the tensor helicities, and the functions $\Gamma_n(\tau)$, for $n\ge 1$,  encode deviations from the Bunch--Davies vacuum. The parameter $\tau_0$ is a reference time introduced to render the expansion dimensionless; it drops out from physical observables.

\paragraph{Iterative solution.}
Following the strategy of~\cite{Tasinato:2023ukp,Tasinato:2023ioq}, we substitute the Ansatz~\eqref{eq_ans2} into Eq.~\eqref{eq:cAk-helicalV2}, expand in powers of $k$, and match coefficients order by order
in $k^n$. This procedure yields a hierarchy of coupled differential equations for the functions $\Gamma_n$, one for each power of $k$. 

At lowest orders, one finds
\begin{align}
\label{condk1}
\partial_\tau \left(\frac{ \tau_0\,\Gamma_1'(\tau)}{\tau^2}  \right)
&=
\partial_\tau \left(\frac{ \gamma' (\tau)}{2\,\tau^2} \right),
\\[6pt]
\label{condk2}
\partial_\tau \left(\frac{\tau_0^2}{\tau^2}  \Gamma_2' (\tau)\right)
&=
\frac{2  \tau_0}{\tau} \partial_\tau \left( \frac{\Gamma_1 (\tau)}{\tau} \right)
+
\frac{\tau}{2}\,\partial_\tau \left(\frac{1}{\tau^2}  \gamma' (\tau)\right),
\end{align}
where primes denote derivatives with respect to conformal time. Higher orders follow a similar recursive structure.

\paragraph{Piecewise evolution.}
The system admits a natural piecewise analysis:
\begin{enumerate}
\item For $\tau<\tau_1$, the source vanishes ($\gamma=0$), implying $\Gamma_n=0$ and recovering the Bunch--Davies solution of Eq.~\eqref{eq_stmf} .
\item For $\tau_1\le \tau \le \tau_2$, the source is active and the full coupled system must be solved.
\item For $\tau>\tau_2$, the system returns to slow roll, and the solution is fixed by matching conditions at $\tau_2$.
\end{enumerate}

\paragraph{Short-duration expansion.}
To make analytical progress during the interval $\tau_1\le \tau \le \tau_2$, we assume that the source is switched on for a short time starting at $\tau_1$, and we 
 impose the conditions 
\begin{equation}
\gamma(\tau_1)=\gamma'(\tau_1)=0,
\qquad
\beta \equiv \frac{\tau_1\,\gamma''(\tau_1)}{2}\,.
\label{def_beta2}
\end{equation}
This choice of these conditions -- compatible with the requirements outlined after Eq.~\eqref{con_short} -- is crucial to carry  our analytical procedure further. 
Taylor
expanding the functions $\Gamma_n$ around $\tau=\tau_1$, and retaining the leading non-vanishing contributions, the hierarchy simplifies. It  admits the following
choice of derivatives as solution at $\tau=\tau_1$
\begin{align}
\tau_0 \Gamma_1^{(2)} (\tau_1)&= \frac{\gamma''(\tau_1)}{2},
\\
\tau_0^2 \Gamma_2^{(2)} (\tau_1)&= \frac{\tau_1\,\gamma''(\tau_1)}{2},
\\
\tau_0^n \Gamma_n^{(n)} (\tau_1)&= 2\,\Gamma_{n-1}^{(n-1)}, \qquad n>2,
\end{align}
where superscripts indicate the order of the derivative. Based on these expressions,
the solution of the system of equations  leads to 
\begin{align}
\tau_0 \Gamma_1 (\tau)&=\frac{ \beta\,( \tau-\tau_1)^2}{\tau_1\,2!},
\\
\tau_0^n \Gamma_n (\tau)&=\frac{ \beta\,2^n\,( \tau-\tau_1)^n}{4\,n!},
\qquad n\ge2
\end{align}
 during the brief interval $\tau_1\le\tau\le\tau_2$.

\paragraph{Resummed solution.}
Substituting back into the Ansatz~\eqref{eq_ans2}, the series can be resummed in a closed form. The original tensor mode function during the transient phase reads
\begin{align}
\label{hk_sol2}
h^{\lambda}_k(\tau)
&=
\frac{H}{\sqrt{2}\,k^{3/2}}\,e^{-i k \tau}
\Bigg[
1 + i k \tau 
+ \frac{i \lambda \beta}{4} \left( 
e^{2 i k (\tau - \tau_1)} 
- 1 
- 2 i k (\tau - \tau_1)\left(1 - \frac{\tau - \tau_1}{\tau_1}\right)
\right)
\Bigg],
\end{align}
valid for $\tau_1\le \tau \le \tau_2$. The parameter $\beta$ of Eq.~\eqref{def_beta2}, proportional to $\gamma''(\tau_1)$, controls the strength of the parity-violating effects and induces helicity-dependent corrections to the mode functions. The structure of 
solution \eqref{hk_sol2} depends  on our Ansatz for the pump field, Eqs.~\eqref{def_pump}
and \eqref{ans_gam}. More general possibilities, motivated by \cite{Crisostomi:2017ugk}, might lead to different results.  

\paragraph{Late-time solution and matching.}
For $\tau\ge \tau_2$, the source switches off again and the solution of Eq.~\eqref{eq:cAk-helicalV2} takes the general form
\begin{equation}
h^{\lambda}_k(\tau)=\frac{H}{\sqrt{2}\,k^{3/2}}\,
\left[ C_1^\lambda\,
e^{-i k \tau}
\left(1+ i k \tau \right)+
 C_2^\lambda\,
e^{i k \tau}
\left(1- i k \tau \right) \right],
\label{eq_stmf2}
\end{equation}
where the time-independent quantities  $C_{1,2}^\lambda$ are determined by matching $h_k$ and $h_k'$ at $\tau=\tau_2$. 
Introducing the dimensionless, positive quantities
\begin{equation}
\label{def_kae}
\kappa \equiv - k \tau_1\,,
\qquad
\Delta \tau \equiv -\frac{\tau_2-\tau_1}{\tau_1}\,,
\end{equation}
the Israel junction conditions dictate
\begin{align}
C_1^\lambda &= 
\frac{1 + \beta\,\lambda\, \left[i - i e^{2 i \Delta\tau \kappa} + 6 \Delta\tau^2 \kappa - 4 i \Delta\tau^3 \kappa^2 + 2 i \Delta\tau (2 + i \kappa + 2 \kappa^2)\right]}{8 (-1 + \Delta\tau)^2 \kappa^2},
\nonumber
\\[8pt]
C_2^\lambda &= 
-\frac{e^{-2 i (-1+\Delta\tau)\kappa} \, \beta \lambda \left[i + 2 \kappa - 2 \Delta\tau^2 \kappa + 4 \Delta\tau (i + \kappa) + e^{2 i \Delta\tau \kappa} \left(-i + 2 (-1+\Delta\tau)\kappa\right)\right]}{8 (-1+\Delta\tau)^2 \kappa^2}\,.
\label{expr_c12}
\end{align}
Notice that the helicity-dependent deviation from the standard Bunch-Davies mode function
is again controlled by the single parameter $\beta$, as defined in Eq.~\eqref{def_beta2}. 
These ingredients allow us to compute the tensor power spectrum in Fourier space at the end inflation, $\tau\to0$, 
 proportional to the square of the mode function $|h_k^\lambda|^2$
for each polarization $\lambda$: 
\begin{eqnarray}
{\cal P}_h^{\lambda}(k)
&=&\frac{k^3}{4 \pi^2}\,|h_k^{\lambda}|^2
\\
&=&
\frac{2 H^2}{\pi^2}\,
\big| C_1^\lambda + C_2^\lambda \big|^2\,.
\label{def_phe}
\end{eqnarray}
The total primordial tensor spectrum -- related with the GW intensity -- is the average 
of the two polarizations:
\begin{eqnarray}
\label{def_opt}
{\cal P}_h^{T}(k)
&=&\frac12 \sum_{\lambda=\pm1} {\cal P}_h^{\lambda}(k)\,.
\end{eqnarray}
At large scales, the spectrum is scale invariant, with an amplitude
 ${\cal P}_h^{T}(\kappa \ll 1) = {2 H^2}/{\pi^2}$. At small
scales, instead, it  results
\be
\label{eq_lark}
{\cal P}_h^{T}(\kappa\gg 1)\,=\,
\frac{2 H^2}{\pi^2}\,\left[
1+\frac{\beta^2 \Delta \tau^2 \,(1+\Delta \tau)^2}{4(1-\Delta \tau)^2}
\right]
\ee
Hence -- in a limit of small duration of PV effects, $\Delta \tau\ll1$ -- we 
find an enhancement of the tensor spectrum from large
towards small scales, controlled by the combination $\beta^2 \Delta \tau^2 $. 
 It is important to stress  the role of helicities in forming the combination
 \eqref{def_opt}. While the terms linear in $\lambda$ cancel, the terms
 quadratic in $\lambda^2$ sum up; hence, new contributions to the GW spectrum 
 towards small scales are associated with quadratic contribution in parity-violating 
 parameters (more on this in Section \ref{sec_pheno}).
 
\smallskip

In  order to investigate the specific scale dependence of the spectrum, 
 it is convenient to compute the dimensionless ratio of the spectrum at momentum $\kappa$
versus its value at $\kappa \to0$  \cite{Tasinato:2020vdk} (the suffix $I$ indicates intensity):
\be
\label{def_pii}
\Pi_I(\kappa)\,\equiv\,\frac{{\cal P}_h^{T}(\kappa)}{{\cal P}_h^{T}(\kappa\to 0)}\,.
\ee
\begin{figure}[t!]
    \centering
    \includegraphics[width=0.49\linewidth]{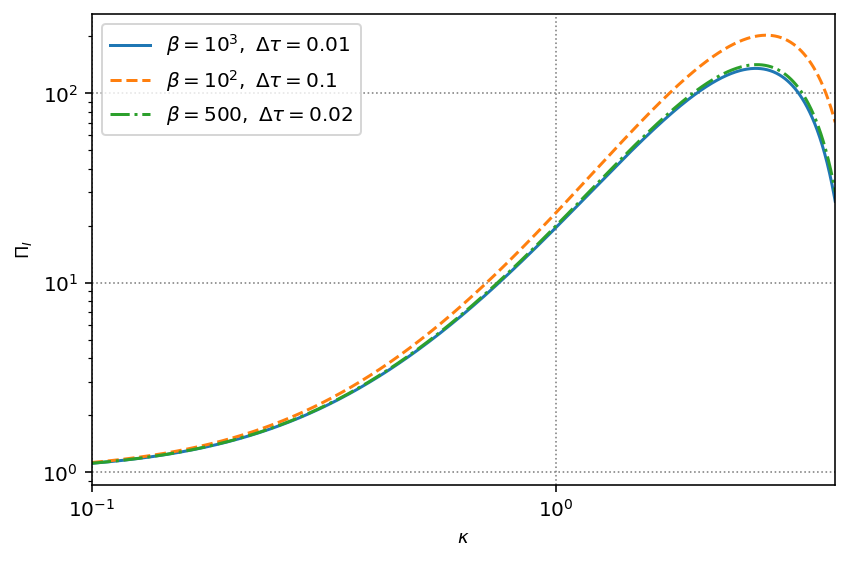}
    \includegraphics[width=0.49\linewidth]{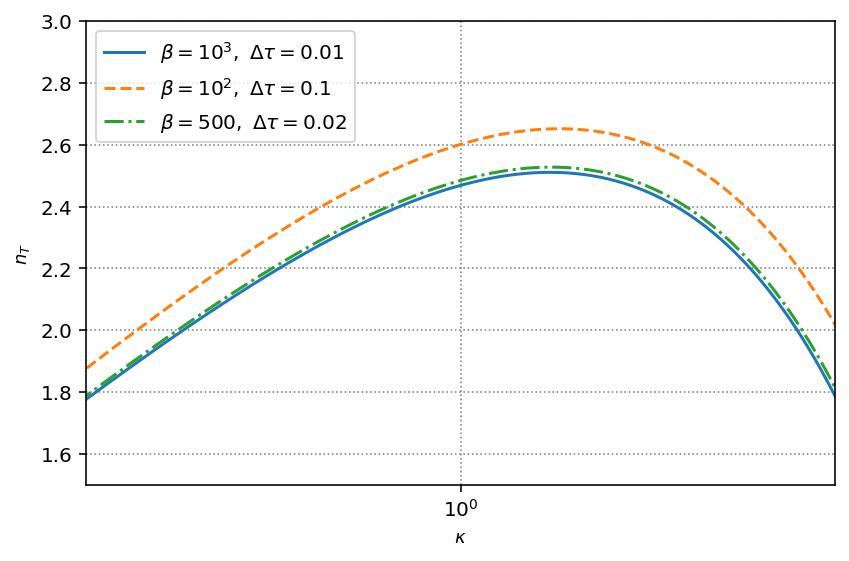}
    \caption{ \small 
  We represent the scale dependence
  of the tensor spectrum generated during inflation on our scenario.
    {\bf Left panel}: plot of Eq.~\eqref{def_pii} in the region of increasing tensor spectrum, from large  towards small scales, for 
    different values of $\beta$ and $\Delta \tau$. {\bf Right panel}: the corresponding spectral index $n_T$ of Eq.~\eqref{def_nT}, zooming around the inflection point of the spectral growth. Notice that, in the growing
    part of the spectrum in proximity of the inflection point, the spectral tilt lies in the range $n_T\simeq 2$.
    }
    \label{fig_spV1}
\end{figure}
 We represent  the quantity $\Pi_I(\kappa)$ in Fig \ref{fig_spV1}, left panel, 
for different values of the parameters  $\beta$ and $\Delta \tau$. On purpose, in all cases we take $\beta$ large and $\Delta \tau$ small, keeping fixed their product. 
After an initial plateau for small $\kappa$ -- which can allow us
to satisfy present bounds on the tensor-to-scalar spectrum at CMB scales, more
on this later -- we notice a steady growth from large 
towards small scales, up to $\kappa \simeq {\cal O}(1)$ where
the spectrum starts to oscillate.  The resulting
plots are  similar   for all cases. In Fig \ref{fig_spV1}, right panel, we instead represent the tensor spectral index during inflation
\be
\label{def_nT}
n_T \,=\,\frac{d \ln \Pi_I}{d \ln \kappa}
\ee
 in the growing region of the spectrum. In all cases, $n_T $ reaches values in the interval around
  $(2, 2.5)$, with a limiting value at around $\sim 2.6$. Hence in our setup  in which PV effects last for a short period of time, we
  find an {\it universal behaviour}  for the growth of GW spectrum during inflation, 
  whose slope reaches  values of considerable phenomenological interest as we discuss in Section
  \ref{sec_pheno}. 
  Our result can be interpreted as the spin-2, parity-violating
analogue of similar bounds found in other sectors. For scalar
perturbations, a brief departure from slow roll leads to a maximal
growth $n_s - 1 = 4$  \cite{Byrnes:2018txb} (up to logarithmic corrections \cite{Carrilho:2019oqg,Ozsoy:2019lyy});
for vector modes one finds $n_V \simeq 4.75$ \cite{Atkins:2025pvg,Ragavendra:2026fgs};
and for the longitudinal mode of a massive dark photon
$n_L \simeq 6$ \cite{Marriott-Best:2025sez,LaRosa:2025woi}.
    Nevertheless, also in the present  case, we interpret the 
  raise of the spectrum as due to the active dynamics of the tensor
  decaying mode of Eq.~\eqref{eq:cAk-helicalV2}, which becomes dominant and amplifies the
  amplitude of the spectrum (see e.g. \cite{Ozsoy:2023ryl} for a review in the context
  of scalar fluctuations). This specific behaviour of the decaying tensor
  mode is induced by the brief, but strong, phase of parity violating interactions.

\paragraph{Large parity-violating expansion.}

For our purposes, we aim to make the duration of the parity-violating (PV) interacting phase as short as possible, $\Delta\tau \ll 1$. This ensures the validity of the Taylor expansion used in determining the mode functions. At the same time, we wish to enhance the inflationary tensor spectrum from large to small scales. 
This requires increasing the amplitude of PV effects, controlled by the parameter $\beta$ (see the small-scale
enhancement of the spectrum, Eq.~\eqref{eq_lark}). 

A convenient and simplifying procedure, inspired by \cite{Tasinato:2023ukp}, is to consider the scaling limit
\begin{equation}
\label{eq_lim2}
\Delta\tau \to 0\,,
\qquad \qquad
\beta \to \infty\,,
\qquad \qquad
\beta\,\Delta\tau =  b_0 = \text{fixed}\,,
\end{equation}
where $b_0$ is a constant parameter. In this limit, large values of $b_0$ lead to a significant enhancement of the tensor spectrum, with the total growth from large to small scales scaling as $b_0^2$ (see Eq.~\eqref{eq_lark}). Such amplification, as mentioned above, is due to the input of the would-be tensor decaying mode, which plays an active role in this set-up. 

The coefficients $C_{1,2}^\lambda$ introduced in Eq.~\eqref{expr_c12} simplify considerably in the limit \eqref{eq_lim2}, reducing to
\begin{align}
C_1^\lambda &= 
1 + \frac{i\,(1+\kappa^2)\,\lambda\,b_0}{2\,\kappa^2}\,,
\label{sim_c1}
\\[6pt]
C_2^\lambda &= 
\frac{i\,e^{2 i \kappa}(i+\kappa)^2\,\lambda\,b_0}{2\,\kappa^2}\,.
\label{sim_c2}
\end{align}
These expressions, once plugged in Eq.~\eqref{eq_stmf2}, make explicit the role of helicity $\lambda$ and the parity-violating parameter $b_0$. We find remarkable that such simple choice of mode functions 
catches well the effects of a transient parity-violating era during inflation.  

The quantity $\Pi_I(\kappa)$ of Eq. \eqref{def_pii}, which controls the scale dependence of the tensor spectrum amplitude, is given by
\begin{equation}
\label{finr_pik}
\Pi_I(\kappa)
=
1 + \frac{b_0^2 (1+\kappa^2)\left(-\kappa \cos\kappa + \sin\kappa\right)^2}{\kappa^4}\,.
\end{equation}
Interestingly, this quantity depends only on two parameters: $b_0$ -- which controls
the amplification of the spectrum from large towards small scales -- and the instant
 $\tau_1$ during inflation -- which in terms of $\kappa = - k \tau_1$ characterizes
 the time at which the short PV epoch occurs. 

\begin{figure}[t!]
    \centering
    \includegraphics[width=0.49\linewidth]{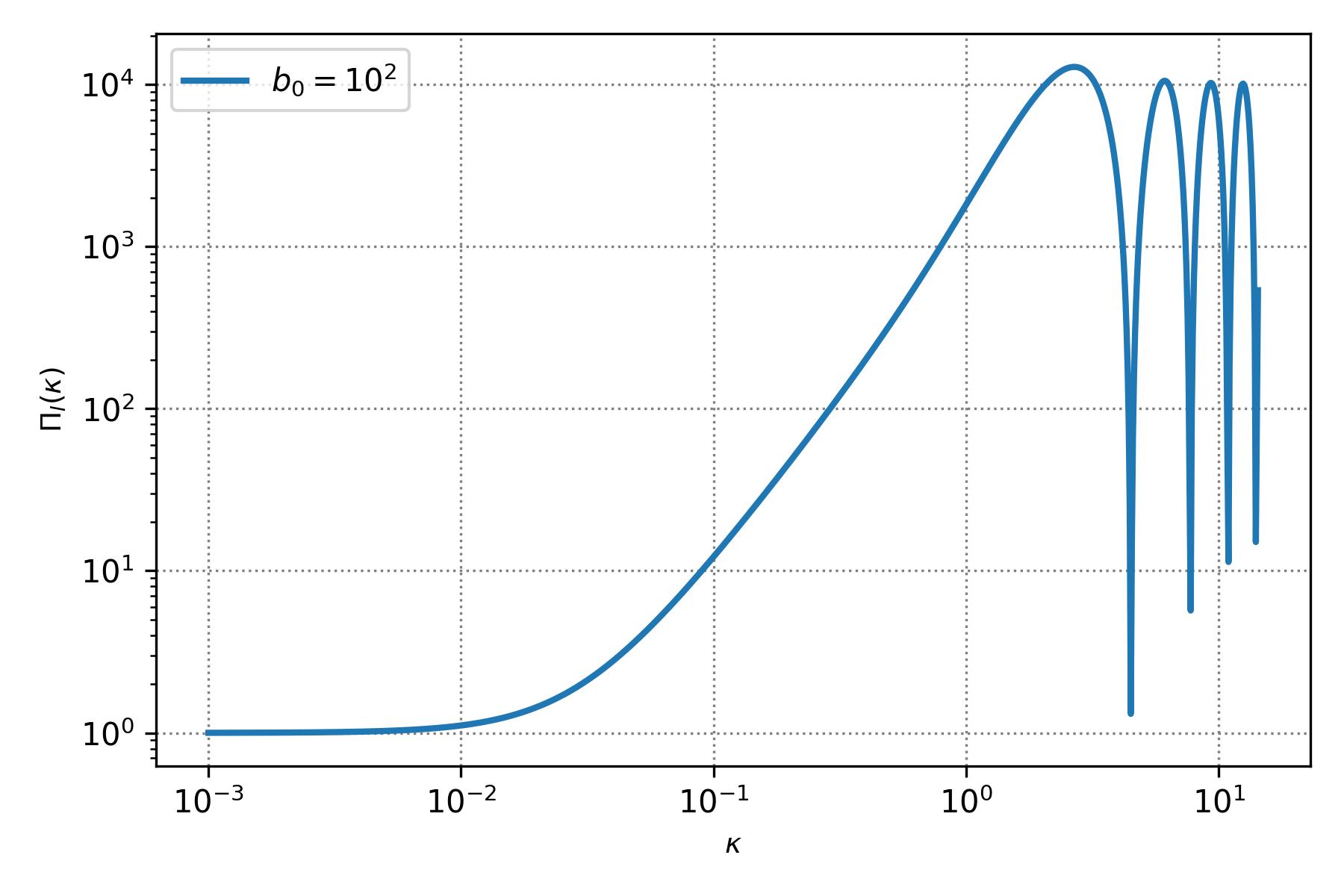}
    \includegraphics[width=0.49\linewidth]{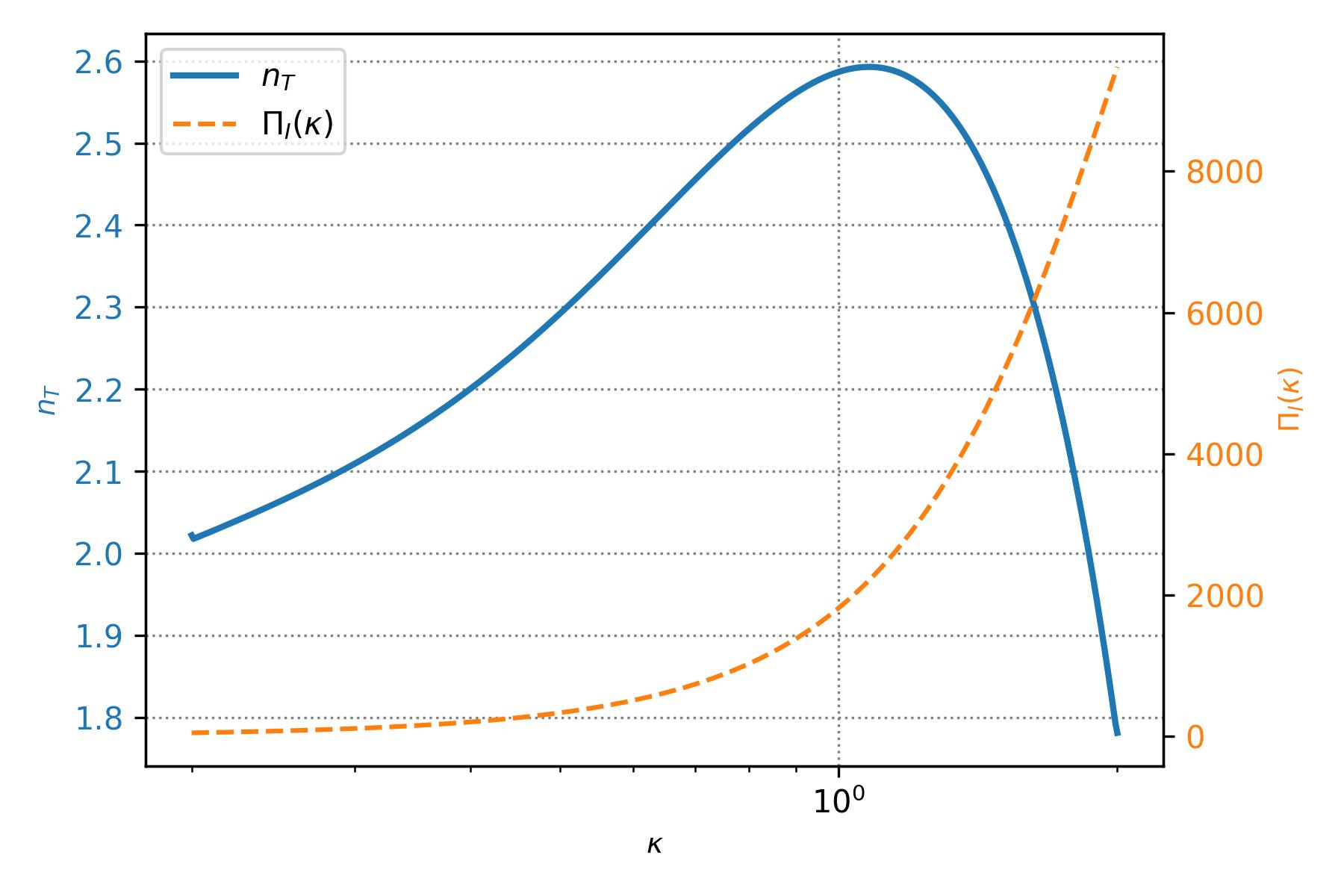}
    \caption{\small {\bf Left panel}: the region of increasing tensor spectrum according
    to Eq.~\eqref{finr_pik}, from large scales (where it is almost scale invariant) towards small scales. {\bf Right panel}: the corresponding spectral index, zooming at the inflection point at around $\kappa\simeq 1$. 
    }
    \label{fig_spV2}
\end{figure}

In Fig.~\ref{fig_spV2}, we show $\Pi_I(\kappa)$ (left panel) together with its slope (right panel). This simplified expression reproduces well the qualitative features of Fig.~\ref{fig_spV1}. In the limit of $b_0$ large (which interests  us)
the spectrum starts and keeps almost scale invariant at large scales, i.e. within the interval
 $0\le \kappa\le \kappa_{\rm growth}$ with
\be
\label{exp_kgr}
\kappa_{\rm growth} \,=\,\frac{3}{b_0}\,,
\ee
where it starts growing with a profile well approximated by a power-law. (We explain the origin of 
Eq.~\eqref{exp_kgr} in Section \ref{sec_pheno}.) Its slope in the growing region 
varies in  a range starting at $n_T\simeq 2$ (recall the definition of $n_T$ in
Eq.~\eqref{def_nT}) with a maximal value $n_T \simeq 2.6$ at around $\kappa = 1$: see Fig.~\ref{fig_spV2}, right panel, where we zoom on the spectral tilt
around the inflection point of the spectrum profile, where the rate of growth
is apparent. 

Phenomenological implications of this scenario, as well as a discussion of gravitational-wave Stokes parameters, are presented in  what comes next.

\section{Phenomenological implications}
\label{sec_pheno}

As discussed in the previous section, a transient phase of
parity violation (PV) during inflation can lead to a significant
enhancement of the tensor spectrum at small scales. Remarkably,
this enhancement can reach amplitudes potentially detectable by
gravitational-wave (GW) observatories such as pulsar timing arrays
(PTAs), while remaining consistent with current constraints on the
tensor-to-scalar ratio at CMB scales. Importantly, it predicts a slope of the
spectrum compatible with recent PTA measurements. 
 In addition, the scenario
predicts distinctive signatures in both the amplitude and the
scale dependence of the GW Stokes parameters. In this section,
we explore these phenomenological implications.

\subsection{Implications for current PTA observations}
\label{sec_pheno_cur}

Starting from the tensor spectrum evaluated at the end of inflation,
$\tau = 0$ (see Eq.~\eqref{def_phe}), we compute the present-day GW energy
density $\Omega_{\rm GW}$ as a function of frequency during the
radiation-dominated era, after tensor modes re-enter the horizon. This quantity is given by
(see e.g.~\cite{Maggiore:2018sht}, Sec.~19.5.4)
\begin{equation}
\label{gw_enden_improved}
h^2\,\Omega^{(I)}_{\rm GW}(f)
\,=\, 6.73\times 10^{-7}\,
{\cal P}_h^{T}\!\left({f}/{f_\star}\right)\,,
\end{equation}
where we used $2\pi f = k$, and defined
\begin{equation}
\label{def_fstar_improved}
2\pi f_\star \,=\, a(\tau_1)\,H\,.
\end{equation}
The superscript $(I)$ denotes the GW intensity, anticipating our later
discussion of the full set of Stokes parameters.
The parameter $\kappa$ introduced in Eq.~\eqref{def_kae} is given by
$\kappa = -k \tau_1 = f/f_\star$. The conformal time $|\tau_1|$
at which the PV phase occurs during inflation sets a characteristic frequency scale
in the GW spectrum.

\paragraph{Frequency profile of $\Omega^{(I)}_{\rm GW}$: general considerations.}

Equation~\eqref{gw_enden_improved} shows that the frequency dependence
of $\Omega_{\rm GW}^{(I)}$ directly traces the scale dependence  of the primordial
tensor spectrum. Therefore, its phenomenology might  be inferred from
the profiles shown in Fig.~\ref{fig_spV2}, expressed in terms of the
dimensionless variable $\kappa$. Accordingly, the spectral tilt $n_T$ corresponds to 
the effective logarithmic slope of the GW energy density,
\begin{equation}
n_T \equiv \frac{d\ln \Omega_{\rm GW}}{d\ln f}\,,
\end{equation}
evaluated over the relevant frequency interval.

At low frequencies, $f/f_\star \ll 1$, the spectrum is approximately
scale invariant up to a characteristic scale
$f_{\rm growth}/f_\star = \kappa_{\rm growth}$ (see Eq.~\eqref{exp_kgr}).
Beyond this point, the spectrum grows steadily, reaching a maximum
around $f/f_\star \sim \mathcal{O}(1)$, and subsequently develops
oscillatory features for $f/f_\star \gtrsim 1$.

In fact, in the intermediate regime  the growth can be well approximated by
a power law --  although the exact profile is not strictly scale-free.
The effective spectral tilt varies in a range starting at $n_T \simeq 2$, with maximal growth occurring near
an inflection point at $f/f_\star \simeq 1$, where
it acquires a  value  $n_T = 2.6$.
In the limit of a short-duration PV phase, this behaviour is
universal and largely insensitive to the microscopic details of the
transition, as illustrated in Fig.~\ref{fig_spV2}. Notably, the range
$n_T \simeq 2$ emerges naturally,
and
overlaps with the effective slopes obtained in PTA analyses when fitting the signal with cosmological power-law templates 
 (see e.g.~\cite{NANOGrav:2023hvm,EPTA:2023xxk,Figueroa:2023zhu,Vagnozzi:2023lwo,Ellis:2023oxs}). 

While the effective slope provides a useful diagnostic, our scenario predicts a richer structure than a simple power law, including a transition from a nearly scale-invariant regime to a growing phase and subsequent oscillatory features. This structure, together with polarization observables we  discuss in Section \ref{sec_stok}, offers a more robust discriminant between different physical origins of the signal.

\paragraph{An explicit example.}

\begin{figure}[t!]
    \centering
    \includegraphics[width=0.49\linewidth]{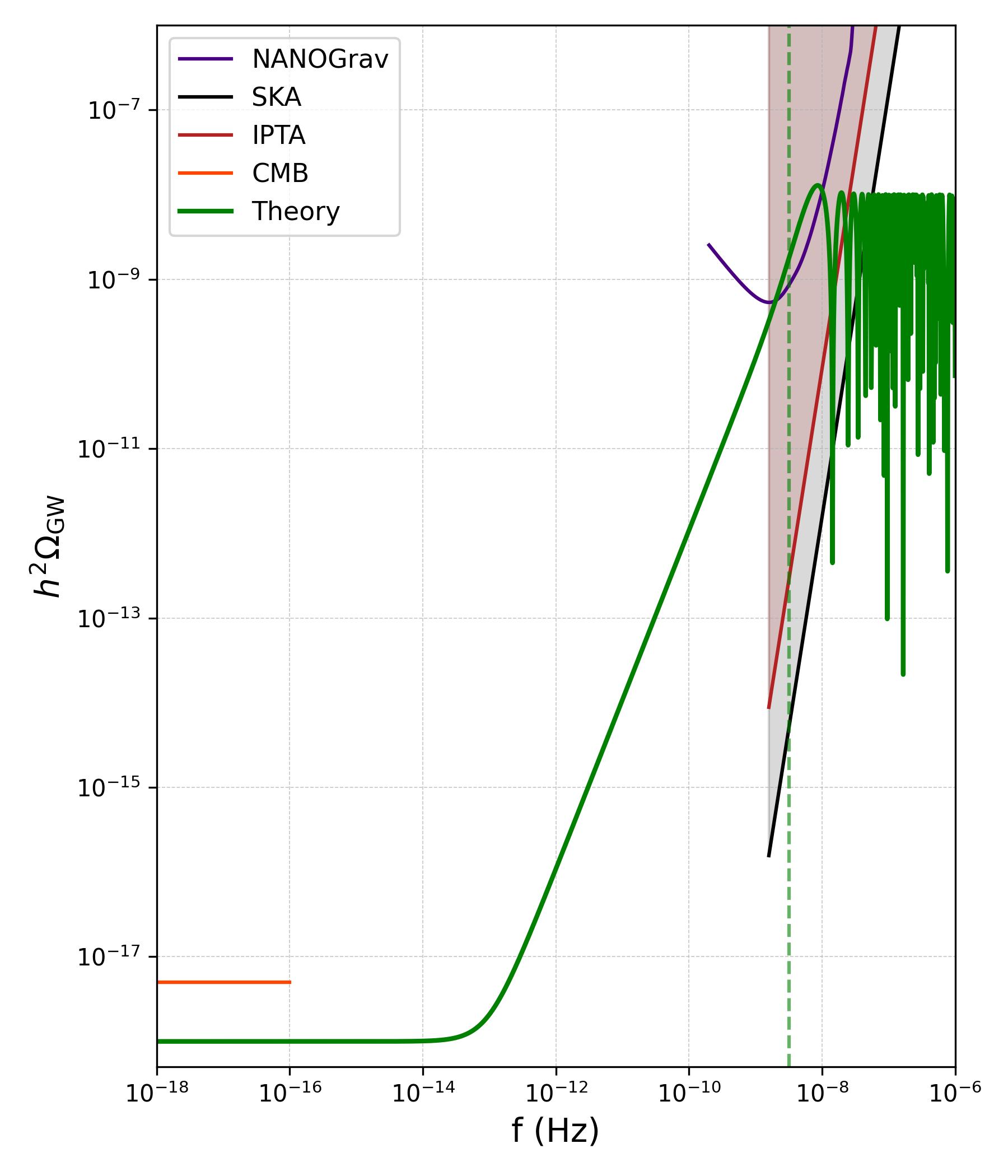}
     \caption{\small 
  Our theoretical prediction for $h^2 \Omega_{\rm GW}^I$ as green line, with parameters discussed after Eq.~\eqref{comp_ogw_improved}, compared with the sensitivity curves of current CMB measurements  \cite{BICEP2:2018kqh}, 
  and of PTA experiments NANOGrav, IPTA, and SKA, taken from
  \cite{Schmitz:2020syl}. 
   }
    \label{fig_spV3}
\end{figure}

We now combine Eqs.~\eqref{finr_pik} and \eqref{gw_enden_improved} to provide a concrete
illustration of the phenomenology outlined above. (For visual reference, see Fig.~\ref{fig_spV3}.) In  our units ($M_{\rm Pl}=1$),
we obtain
\begin{equation}
\label{comp_ogw_improved}
h^2\,
\Omega_{\rm GW}^{(I)}(f) \,=\, 2.73\times 10^{-7}\, H^2\,
\left\{
1 + \frac{b_0^2 (f_\star^2+f^2)\left[-f \cos\!\left({f}/{f_\star}\right)
+ f_\star \sin\!\left({f}/{f_\star}\right) \right]^2}{f^4}
\right\} .
\end{equation}

At large scales (CMB frequencies), the spectrum is approximately
scale invariant. This sets a bound on the Hubble parameter $H$ during inflation. Using the approximate  relation
$h^2 \Omega_{\rm GW}^{(I)} \sim 10^{-15} r$ \cite{Maggiore:2018sht},
we fix the amplitude by choosing $r = 10^{-3}$, which corresponds to
$h^2 \Omega_{\rm GW}^{(I)}(f \ll f_\star) \simeq 10^{-18}$ and
$H = 3 \times 10^{-6}$. Such values of $r$ are within reach of
next-generation CMB B-mode experiments \cite{LiteBIRD:2022cnt,CMB-S4:2022ght} -- hence
in our setup  primordial tensor modes might
be detected also at  CMB scales.

\smallskip

At larger nanohertz frequencies, current PTA observations \cite{NANOGrav:2023gor,Reardon:2023gzh,Xu:2023wog,EPTA:2023fyk} provide evidence
for a stochastic GW background with significantly larger amplitude.
We can use PTA measurements to set our remaining free parameters $b_0$, $f_\star$. 
Adopting a power-law parameterization for the GW energy density in the PTA band,
$\Omega_{\rm GW}^{(I)}(f) = A_\star (f/f_{\rm yr})^{n_T}$,
Ref.~\cite{Figueroa:2023zhu} report the  values
\be
(A_\star , n_T) =
\left(4.25^{+3.02}_{-1.92} \times 10^{-8},\;
2.08^{+0.32}_{-0.30}\right) \,.
\ee
In our framework, these values can be reproduced by fixing $b_0$ to match the amplitude of the spectrum at small scales (which scales as $b_0^2$), and $f_\star$ to determine the frequency at which the spectrum begins to grow (see Fig.~\ref{fig_spV3}). 
Accordingly, we adopt $b_0 = 10^{5}$ and $f_\star = 3.2 \times 10^{-9}\,\mathrm{Hz}$ in Eq.~\eqref{comp_ogw_improved}. 
Importantly, the spectral slope is \textit{not} a free parameter; rather, our choice ensures that the (approximate) GW power-law spectrum crosses the PTA band, with its predicted slope $n_T \simeq 2$. 
As a result, the GW spectrum remains consistent with current CMB bounds on large scales, while naturally developing the amplitude and spectral tilt compatible with
  PTA observations at smaller scales.\footnote{For this value of $f_\star$, Eq.~\eqref{exp_kgr} implies that the GW spectrum starts to grow only at frequencies $f_{\rm growth} \simeq 10^{-13}\,\mathrm{Hz}$, well outside the range probed by CMB polarization experiments.}
In Fig.~\ref{fig_spV3}, we compare our theoretical predictions with the sensitivity curves of current and future GW experiments in the nano-Hertz band. 

\smallskip

While our conclusions are supported by our qualitative arguments and previous studies, it would be important to assess quantitatively and more precisely how well the GW energy density in Eq.~\eqref{comp_ogw_improved}, characterized by an approximate power-law behavior in the PTA band, fits recent PTA data. 
We defer this analysis to future work.

\subsection{New predictions: GW Stokes parameters}
\label{sec_stok}

Interestingly, our scenario makes further predictions which can provide important
 signatures to prove (or disprove) our ideas. 
 The two-point correlators of GW, which are helicity-two quantities, can be decomposed
 in terms of Stokes parameters, analogously to electromagnetic fields. They
 are  intensity $I$, circular polarization $V$, and linear polarizations $U$ and $Q$.
 See e.g. \cite{Smith:2016jqs} for a nice discussion
on Stokes parameters for GW backgrounds.
 In terms of mode functions $h^{L,R}$ of eqs \eqref{eq_stmf2} evaluated the end of inflation and expressed in circular basis, they read
 \begin{equation}
\begin{aligned}
I &= \frac12 \left( |h^L|^2 + |h^R|^2\right), \hskip1.3cm
V = \frac12 \left( |h^R|^2 - |h^L|^2\right),\\
Q  &= \, \mathrm{Re}\big(h^R (h^L)^*\big), \hskip2.5cm
U = -\, \mathrm{Im}\big(h^R (h^L)^*\big).
\end{aligned}
\end{equation}
Calling $X=I,\dots U$, 
 we define  the quantities $\Pi_X = X(\kappa)/I(\kappa\to 0)$ which control the momentum dependence
 of Stokes parameters for inflationary  GW. Using Eqs.~\eqref{sim_c1} and \eqref{sim_c2},  
we find for our setup 
\begin{eqnarray}
\Pi_I &=& 1 + \frac{b_0^2 (1+\kappa^2)\left(-\kappa \cos\kappa + \sin\kappa\right)^2}{\kappa^4} \,,\hskip0.4 cm
\Pi_V \,=\, \frac{b_0 \left(2 \kappa \cos(2\kappa) + (-1+\kappa^2)\sin(2\kappa)\right)}{\kappa^2} 
\nonumber
\\
\Pi_Q &=& 1 - \frac{b_0^2 (1+\kappa^2)\left(-\kappa \cos\kappa + \sin\kappa\right)^2}{\kappa^4} \,,\hskip0.4 cm
\Pi_U \,=\, \frac{2 b_0 \left(-\kappa \cos\kappa + \sin\kappa\right)^2}{\kappa^2}\,.
\label{eq_stpar}
\end{eqnarray}
\begin{figure}[t!]
    \centering
    \includegraphics[width=0.6\linewidth]{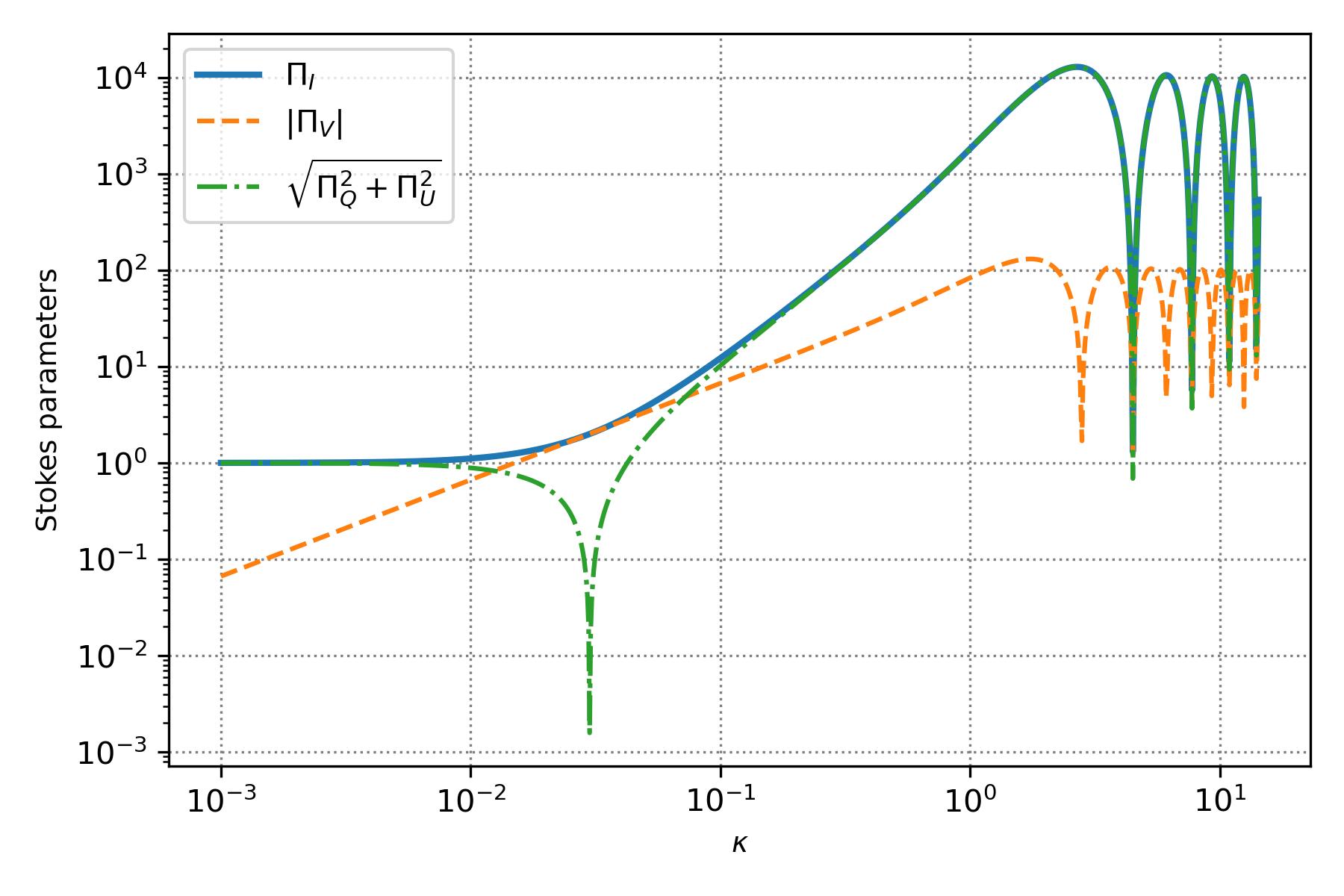}
     \caption{\small 
The functions defined in Eq.~\eqref{eq_stpar} describing the scale dependence of the GW Stokes parameters in our setup. While circular polarization is generated, it remains suppressed, whereas the linear polarization reaches a sizeable amplitude comparable to intensity.
   }
    \label{fig_spV4}
\end{figure}
 The consequences of these results can be summarized as follows. (See Fig.~\ref{fig_spV4}.)

\paragraph{Circular polarization.}
The parity-violating nature of the Chern--Simons--type couplings considered in this work generically induces a non-vanishing circular polarization. Its amplitude is controlled by a single power of the parameter $b_0$, see Eq.~\eqref{eq_stpar}, while intensity scales as $b_0^2$. Adopting the representative value of $b_0$ discussed in Section~\ref{sec_pheno_cur}, the maximal amplitude of circular polarization is expected to be significantly suppressed with respect to the intensity, e.g.\ by a factor $1/b_0 \sim 10^{-5}$. 

Such a suppression makes circular polarization a challenging observable for PTA measurements~\cite{Kato:2015bye}, unless one exploits parity-odd anisotropies (see e.g~\cite{Tasinato:2023zcg,Cruz:2024esk}
and references therein). Despite this limitation, the behaviour of $|\Pi_V|$, shown in Fig.~\ref{fig_spV4}, exhibits noteworthy features. In particular, $|\Pi_V|$ scales linearly with $\kappa$, and reaches the same order of magnitude as $\Pi_I$ precisely at the scale where $\Pi_I$ transitions from an approximately constant regime to a phase of growth. 

We have verified that this transition occurs at the value of $\kappa$ for which the mode function $h^R_k$ vanishes when evaluated at the end of inflation, $\tau = 0$. This characteristic scale coincides with $\kappa_{\rm growth}$ introduced in Eq.~\eqref{exp_kgr}, and, in the large-$b_0$ limit, can be determined analytically using the method developed in~\cite{Tasinato:2020vdk} (see in particular its Section~3), by Taylor expanding the argument
of $h^R_k$ in powers of $1/b_0$.

\paragraph{Linear polarization.}
The linear polarization parameters $Q$ and $U$ are not scalar quantities, as they transform under rotations of the polarization angle. However, the combination $\Pi_L \equiv \sqrt{\Pi_Q^2 + \Pi_U^2}$ is rotationally invariant, and we display it in Fig.~\ref{fig_spV4}. 
Remarkably, in the present setup this quantity can become large, with both its amplitude and scale dependence closely tracking those of the intensity at small scales. From Eqs.~\eqref{eq_stpar}, this behavior originates from the fact that $\Pi_I$ and $\Pi_Q$ scale quadratically with the parameter $b_0$, while $\Pi_U$ is only linear in $b_0$. 

As a result, in the PTA band the linear polarization fraction satisfies
\begin{equation}
\Pi_L \simeq \Pi_I \, ,
\end{equation}
up to small corrections controlled by $\Pi_V/\Pi_I$. This implies that the stochastic background is nearly maximally linearly polarized at the scales of interest.
Such a large linear polarization signal would constitute a striking observational signature of our scenario, especially in view of the fact that typical astrophysical sources are expected to generate only a small degree of linear polarization, at the level of a few percent of the total intensity~\cite{Sato-Polito:2023spo}. If detectable---for instance, with techniques such as those proposed in~\cite{Chu:2021krj,AnilKumar:2023hza}---this feature would provide a strong indication of a cosmological origin of the signal. Hence dedicated forecasts to assess the sensitivity of forthcoming PTA experiments to such observables would therefore be  desirable. And besides experimental consequences, such a large linear polarization has interesting theoretical ramifications too, as we are going to discuss.

\section{Polarization and coherence of the GW signal}
\label{sec_quantum}

An interesting consequence of the  results of Section~\ref{sec_stok} is that PTA observations may provide some hints not only on the primordial origin of the stochastic background, but also on the quantum coherence properties of the underlying tensor fluctuations. Quantum
aspects of primordial tensor modes from inflation is an interesting and
well studied subject, starting from \cite{Grishchuk:1989ss,Grishchuk:1990bj}.
 Interestingly, 
 in our setup  the very same transient parity-violating phase that amplifies the GW intensity also induces a highly correlated polarization pattern -- which can provide suggestive evidence for quantum effects.

\medskip

Recall that GW linear polarization is controlled by the helicity-space correlator
\begin{equation}
Q+iU = h^R (h^L)^* \, ,
\end{equation}
which depends on the relative phase between the two helicities. In contrast to the intensity, which probes only $|h^{L,R}|^2$, the quantities $Q$ and $U$ are directly sensitive to phase coherence among modes of different
helicity. 

The explicit expressions in Eq.~\eqref{eq_stpar} lead to  the identity
\begin{equation}
\Pi_I^2 = \Pi_V^2 + \Pi_Q^2 + \Pi_U^2 \, ,
\end{equation}
corresponding to the saturation of the positivity bound for the polarization matrix \cite{Smith:2016jqs}. Hence the polarization matrix has rank one: the two helicities are  not independent stochastic variables but are 
 strongly phase-correlated.
 Consequently, the stochastic background is nearly maximally polarized, with its  linear polarization tracking the intensity at PTA scales
(in fact, see Fig.~\ref{fig_spV4}).
This is a non-generic property for stochastic GW backgrounds. In fact, in classical stochastic backgrounds generated by incoherent superpositions of many independent sources, phase correlations between helicities are expected to average out, leading to suppressed linear polarization. See e.g. the explicit calculation of \cite{Sato-Polito:2023spo} for astrophysical
sources in the PTA band. 
 Reproducing  a configuration like this  -- although not impossible --   would  require  a remarkably finely tuned classical ensemble, with phase locking between $h_R$ and $h_L$ for each Fourier mode.

\smallskip

We can explicitly trace
the origin of this behavior in our setup, by expressing some of our results
  by means of   Bogoliubov coefficients. Schematically, the late-time mode functions towards the end 
  of inflation  can be written as
\begin{equation}
h_k^\lambda = \alpha_k^\lambda\,u_k + \beta_k^\lambda\,u_k^* \, ,
\end{equation}
with $\alpha_k^\lambda=C_1^\lambda$ and $\beta_k^\lambda=C_2^\lambda$. 
Notice that, correctly, $|\alpha_k^\lambda|^2-|\beta_k^\lambda|^2=1$. 
Using Eqs.~\eqref{sim_c1} and \eqref{sim_c2}, one finds that the occupation number with respect to the Bunch--Davies basis is
(see e.g. \cite{Maggiore:2018sht})
\begin{equation}
n_k^\lambda = |\beta_k^\lambda|^2 = \frac{b_0^2(1+\kappa^2)^2}{4\kappa^4} \, ,
\end{equation}
which is identical for the two helicities:  the large polarization effects found in our model are {\it not} associated with different occupation numbers for left- and right-handed modes. 

\smallskip

The crucial quantity to examine is instead the phase-sensitive combination of Bogoliubov coefficients. Defining
\begin{equation}
\phi_k^\lambda \equiv \frac12\,\arg\!\left(\frac{\beta_k^\lambda}{\alpha_k^\lambda}\right),
\end{equation}
we obtain, using the results of Section~\ref{sec_setup},
\begin{equation}
\phi_k^\lambda
=
\kappa+\arctan\!\left({\kappa}^{-1}\right)
+\frac{\lambda\pi}{4}
-\frac12\arctan\!\left(\lambda\,\frac{(1+\kappa^2)b_0}{2\kappa^2}\right),
\qquad {\rm mod}\,\pi \, .
\label{eq_phase_exact}
\end{equation}
In the phenomenologically relevant regime of large $b_0$, this expression reduces to
\begin{equation}
\phi_k^\lambda
\simeq
\kappa+\arctan\!\left({\kappa}^{-1}\right)
+\lambda\,\frac{\kappa^2}{b_0(1+\kappa^2)},
\qquad {\rm mod}\,\pi \, .
\label{eq_phase_approx}
\end{equation}
These expressions clarify the physical origin of the effect.
 On one hand,
the transient parity-violating phase  generates a large excitation above the vacuum, quantified by the occupation number $n_k^\lambda$ which is proportional
to $b_0^2$ (for us a large quantity).
 On the other hand, it imprints a non-vanishing helicity-dependent phase in the Bogoliubov coefficients. It is precisely this phase structure that controls the interference between positive- and negative-frequency components, and therefore sources the large correlator $h^R (h^L)^*$ responsible for linear polarization. In other words, the GW intensity is mainly controlled by the amount of excitation, while the linear polarization  profile probes the coherence pattern encoded in the squeezing phase.

Hence, the large value of linear polarization $\Pi_L$ in our setup -- together with its scale dependence -- does not arise because one helicity is produced more efficiently than the other, but because the transient parity-violating era fixes a definite relative phase between the two. The resulting polarization pattern is therefore a direct manifestation of coherent mode mixing.
While a classical stochastic description might in principle reproduce these results, the specific combination of features predicted in our scenario is highly non-trivial: a large occupation number generated by vacuum amplification, acting in synergy with a definite phase relation between helicities leading to an almost maximally polarized state.

The simultaneous observation of a growing spectrum with $n_T \simeq 2$ and such a strong, coherent linear polarization would thus provide circumstantial evidence for a primordial, coherently generated origin of the signal, as naturally expected from the amplification of quantum vacuum fluctuations during inflation, although not uniquely excluding more contrived classical scenarios. 
 The study of further observables, for example associated with higher order cosmological 
correlators~\cite{Ciprini:2026pvz}, would strengthen these results and deserves to be investigated. 

\section{Outlook}
\label{sec_out}

The origin of the stochastic gravitational-wave background remains underdetermined, with both astrophysical and cosmological interpretations currently viable. In this context, it is essential to develop predictive theoretical frameworks with minimal parametric freedom and distinctive observational signatures. The scenario presented in this work contributes to this effort by providing a concrete cosmological template characterized by a universal spectral profile and non-trivial polarization structure. In synergy, such features  offer new avenues to disentangle primordial and astrophysical contributions to the signal as observational capabilities improve.

\smallskip

Specifically, we have proposed a scenario in which a transient phase of parity-violating interactions amplifies the spectrum of inflationary gravitational waves at small scales. The resulting stochastic background exhibits robust features in frequency space, most notably an approximate power-law behaviour with a slope compatible with that inferred from recent PTA observations. In addition, the framework predicts distinctive polarization signatures, including a sizeable linear polarization component, while circular polarization remains subdominant. If detected, these features would provide suggestive indications of a primordial, coherently generated origin of the signal.

The perspective opened by our results is twofold, bridging observational opportunities and theoretical developments.

\noindent
{\bf{Experiment:}}
On the observational side, a natural next step is a systematic comparison of the predicted GW spectrum with current and forthcoming PTA datasets, aiming at a global fit of the model parameters and a quantitative assessment of its approximate power-law behaviour at nano-Hertz scales. It will also be important to develop dedicated pipelines to probe the polarization content of the stochastic background. In this context, forecasts for the amplitude and scale dependence of circular polarization could clarify whether such a signal lies within reach of next-generation PTA analyses, while the sizeable linear polarization predicted here may provide a powerful discriminator of the scenario.

\smallskip
\noindent
{\bf{Theory:}}
On the theoretical side, our construction suggests the presence of an underlying mechanism responsible for the apparent universality of the spectral slope. Elucidating the physical origin of this behaviour—possibly linked to the dynamics of a would-be decaying mode that becomes operative during the parity-violating phase—represents an important direction for future work, and may benefit from an effective field theory formulation. 
It will also be valuable to construct explicit realizations of the Ansatz \eqref{ans_gam}, assess their theoretical consistency (including the absence of ghost-like instabilities in Chern--Simons--type gravity), and explore their observational implications. 
Furthermore, extending the analysis to higher-order correlation functions is particularly promising: non-trivial signatures, including violations of standard consistency conditions in squeezed limits due to the would-be decaying tensor mode, could arise and lead to sharp, testable predictions, offering additional probes of the origin of the GW background.

\smallskip

Taken together, these directions point to a rich phenomenology and a set of theoretical questions that can be explored with upcoming data and improved theoretical tools. Work along these lines is currently in progress, and we expect it to further clarify the observational viability and fundamental implications of the scenario proposed here.

\subsection*{Acknowledgments}

It is a pleasure to thank Laura Iacconi for important discussions and useful remarks. 
Interesting discussions with, and input from,   Ameek Malhotra, Alisha Marriott-Best, and Maria Mylova are also gratefully acknowledged. This research is partially
funded by the STFC grants ST/T000813/1 and ST/X000648/1.

  {\small
  
  \providecommand{\href}[2]{#2}\begingroup\raggedright\endgroup

}

\end{document}